\documentclass[11pt,twoside]{article}

\usepackage{asp2021}

\aspSuppressVolSlug
\resetcounters

\begin{document}

\title{Teareduce: a Python package with utilities for teaching reduction
techniques in Astronomy}

\author{Nicol\'{a}s Cardiel,$^{1,2}$ 
Sergio Pascual,$^{1,2}$ 
Mar\'{\i}a Chillar\'{o}n-V\'{\i}ctor,$^{1,2}$ 
Cristina Cabello,$^{1,2}$ 
Jes\'{u}s Gallego,$^{1,2}$
Jaime Zamorano,$^{1}$
and 
Mar\'{\i}a Teresa Ceballos$^{3}$}
\affil{$^1$Departamento de F\'{\i}sica de la Tierra y Astrof\'{\i}sica,
Facultad CC.~F\'{\i}sicas, Universidad Complutense de Madrid, E-28040 Madrid, Spain; 
\email{cardiel@ucm.es}}
\affil{$^2$Instituto de F\'{\i}sica de Part\'{\i}culas y del Cosmos, IPARCOS,
Facultad CC.~F\'{\i}sicas, Universidad Complutense de Madrid, E-28040 Madrid, Spain}
\affil{$^3$Instituto de F\'{\i}sica de Cantabria, CSIC-Universidad de
Cantabria, E-39005 Santander, Spain}

\paperauthor{Nicol\'{a}s Cardiel}{cardiel@ucm.es}{0000-0002-9334-2979}{Universidad Complutense de Madrid}{F\'{\i}sica de la Tierra y Astrof\'{\i}sica}{Madrid}{Madrid}{28040}{Spain}
\paperauthor{Sergio Pascual}{sergiopr@fis.ucm.es}{0000-0002-9351-6051}{Universidad Complutense de Madrid}{F\'{\i}sica de la Tierra y Astrof\'{\i}sica}{Madrid}{Madrid}{28040}{Spain}
\paperauthor{Mar\'{\i}a Chillar\'{o}n-V\'{\i}ctor}{mchill01@ucm.es}{0009-0001-4118-6927}{Universidad Complutense de Madrid}{F\'{\i}sica de la Tierra y Astrof\'{\i}sica}{Madrid}{Madrid}{28040}{Spain}
\paperauthor{Cristina Cabello}{criscabe@ucm.es}{0000-0003-4187-7055}{Universidad Complutense de Madrid}{F\'{\i}sica de la Tierra y Astrof\'{\i}sica}{Madrid}{Madrid}{28040}{Spain}
\paperauthor{Jes\'{u}s Gallego}{j.gallego@ucm.es}{0000-0003-1439-7697}{Universidad Complutense de Madrid}{F\'{\i}sica de la Tierra y Astrof\'{\i}sica}{Madrid}{Madrid}{28040}{Spain}
\paperauthor{Jaime Zamorano}{jzamorano@fis.ucm.es}{0000-0002-8993-5894}{Universidad Complutense de Madrid}{F\'{\i}sica de la Tierra y Astrof\'{\i}sica}{Madrid}{Madrid}{28040}{Spain}
\paperauthor{Mar\'{i}a Teresa Ceballos}{ceballos@ifca.unican.es}{0000-0001-6074-3621}{Instituto de F\'{i}sica de Cantabria}{}{Santander}{Cantabria}{39005}{Spain}



\begin{abstract}
The Python package \texttt{teareduce} has been developed to support teaching
activities related to the reduction of astronomical data. Specifically, it
serves as instructional material for students participating in practical
classes on the processing of astronomical images acquired with various
instruments and telescopes. These classes are part of the course Experimental
Techniques in Astrophysics, which belongs to the Master's Degree in
Astrophysics at the Complutense University of Madrid. The code is publicly
available on GitHub, accompanied by a documentation page that includes Jupyter
notebooks demonstrating the use of its various classes and functions.
\end{abstract}



\section{Introduction}

Python is widely used in the scientific community and offers a powerful
ecosystem for astronomical data processing. \texttt{Astropy} and affiliated
packages provide ready-to-use tools for tasks like handling FITS files,
performing calibrations, and analyzing photometric data, which significantly
simplifies the image reduction workflow. Additionally, Python integrates
seamlessly with other scientific packages (e.g., \texttt{NumPy},
\texttt{SciPy}, \texttt{Matplotlib}) to enhance data manipulation and
visualization. Its cross-platform compatibility with Windows, macOS, and Linux
further ensures flexibility and accessibility for researchers working in
different computing environments, making Python an ideal choice for
astronomical image reduction.

\section{Teaching reduction techniques in the classroom}

As part of the course \textit{T\'{e}cnicas Experimentales en Astrof\'{\i}sica}
(TEA; Experimental Techniques in Astrophysics), included in the Master's Degree
in Astrophysics at the Universidad Complutense de Madrid (Spain), students
prepare observing proposals, carry out astronomical observations using the
CAFOS instrument installed on the 2.2 m telescope at the Calar Alto Observatory
(Almer\'{\i}a, Spain; see Fig.~\ref{fig1}), reduce the acquired images, and
analyse the results. 

\articlefigure[width=1.00\textwidth]{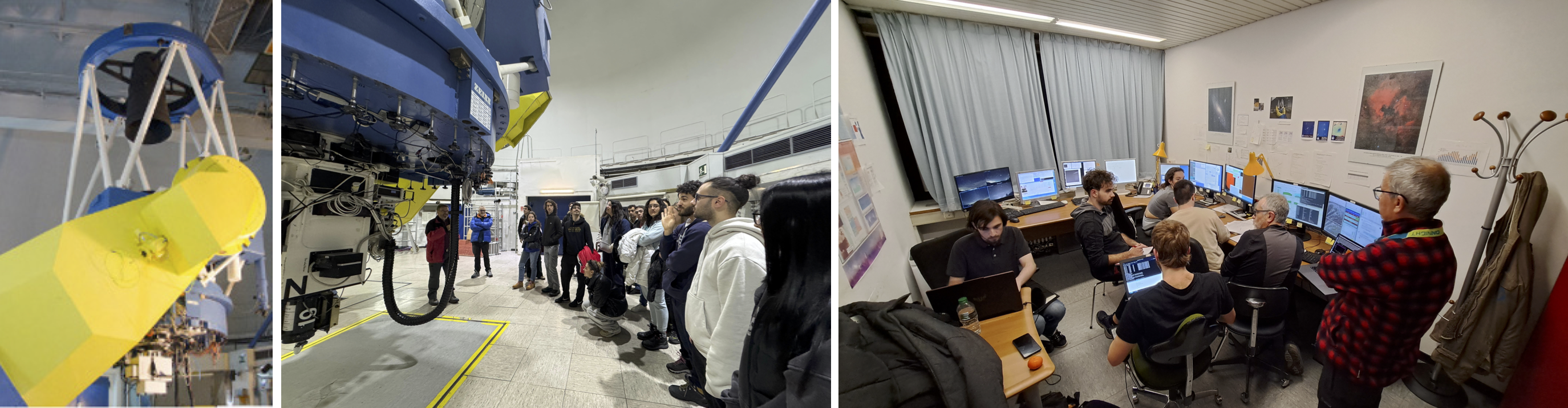}{fig1}{\textit{Left:} the 2.2m Telescope at the
Calar Alto Observatory. \textit{Center:} students listening to explanations
from the observatory staff on the use of the CAFOS instrument.
\textit{Right:} students and faculty of the Master's program conducting
observations from the control room of the 2.2m Telescope.}

To support the data reduction and analysis process, we have developed the
Python package \texttt{teareduce}\footnote{\url{https://nicocardiel.github.io/teareduce-cookbook/intro.html}}.
Its main goal of is to serve as an educational tool.

\texttt{Teareduce} is not intended as a general-purpose image reduction tool.
Rather, it includes a set of specific operations needed for certain steps in
the traditional astronomical image reduction workflow. At the time of its
creation, these features were not available in more established packages such
as \texttt{ccdproc}. Students can examine the Python code and introduce modifications in
order to gain a deeper understanding of the operations performed during the
astronomical image reduction process. The package also introduces alternative
approaches to some tasks, which have proven to be more practical for use in
Master's-level instruction. Some examples are shown below.

\subsection{Auxiliary classes for image slicing}

Working with astronomical images in FITS (Flexible Image Transport System)
format is a fundamental task in modern astronomy. However, one of the most
common and persistent sources of programming errors when handling FITS files in
Python stems from the different indexing and slicing conventions used across
various systems. These differences, rooted in historical computing decisions,
continue to confuse astronomers and programmers alike, leading to subtle bugs
that can compromise scientific results.

The \texttt{teareduce} classes
\texttt{SliceRegion1D}, \texttt{SliceRegion2D}, and \texttt{SliceRegion3D}
allow the users to explicitly define the slice region following either FITS or
Python indexing convention. Once an object is instantiated, it can also be used
with either the FITS or Python convention by simply accessing the
corresponding attribute: \texttt{.fits} or \texttt{.python}.

\subsection{Wavelength calibration}

The C-distortion correction and the wavelength calibration of astronomical
spectra are very important steps in the reduction of astronomical spectra. The
\texttt{teareduce} package provides the \texttt{TeaWaveCalibration} class,
which enables the following steps to be performed: interactive identification
of arc lines, automatic detection of line peaks across all spectra in a 2D
image, and computation of polynomial corrections to perform the C-distortion
correction and the wavelength calibration (see Fig.~\ref{fig2}). The resulting
calibration can be applied both to \texttt{Numpy} arrays and to
\texttt{CCDData}\footnote{\url{https://docs.astropy.org/en/stable/nddata/ccddata.html}} objects.

\articlefigure[width=1.00\textwidth]{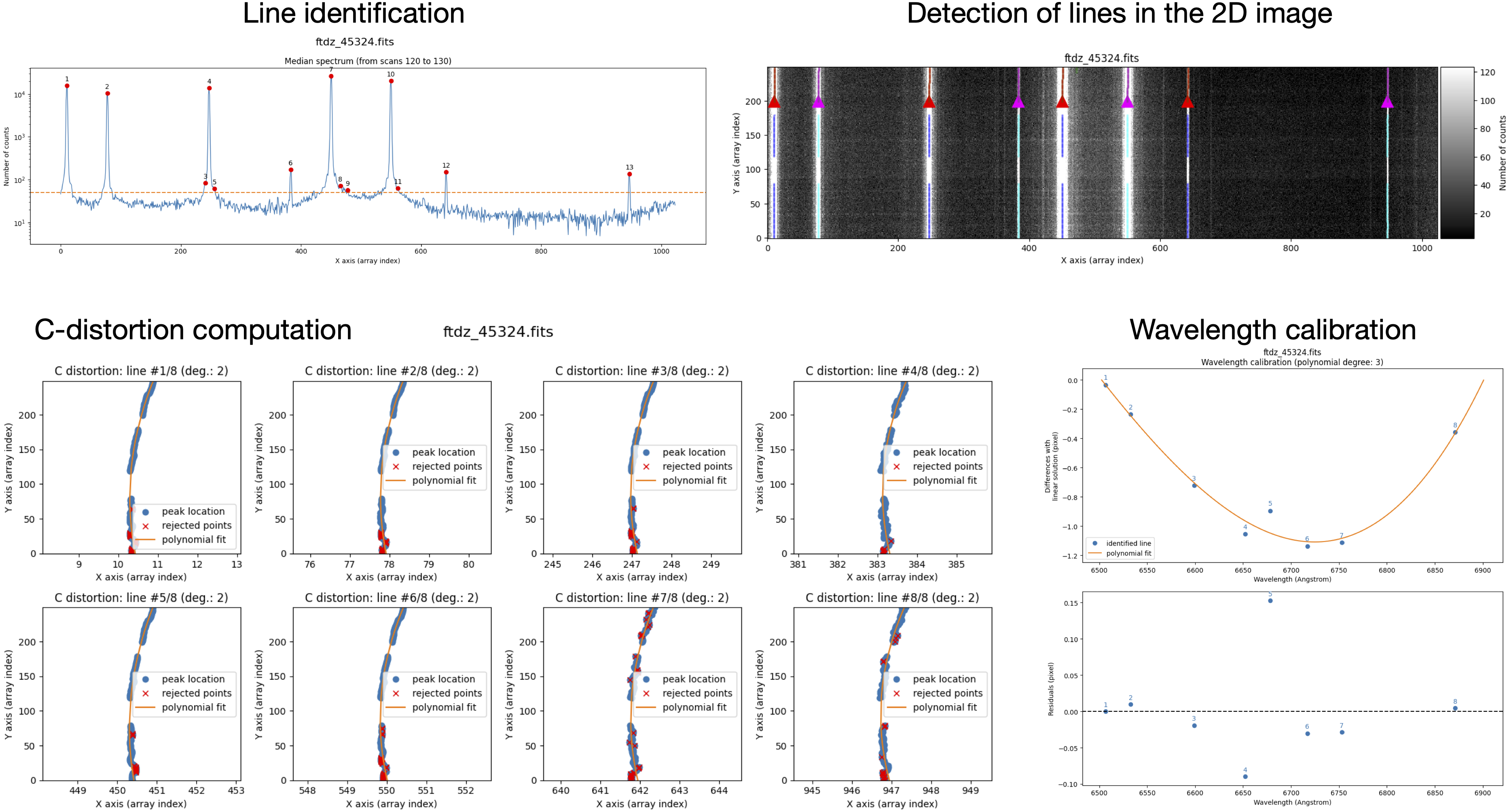}{fig2}{Different steps carried out
by the \texttt{TeaWaveCalibration} class.}

\subsection{Adaptive spline fit}

There are many situations in which it is useful to perform a smooth fit to a
series of points, but a polynomial fit does not provide the necessary
flexibility. In such cases, spline fitting is often used. 

To facilitate this task, the class \texttt{AdaptiveLSQUnivariateSpline} allows
performing this type of fit without having to predefine the locations of the
knots (the points where the different polynomial segments join to create a
continuous and smooth curve), requiring only the specification of the number of
intermediate knots to use \citep[see e.g.,][]{Cardiel2009}. Applying a
numerical minimisation process allows the fit to automatically adjust the knot
positions in order to achieve a good match to the data.

\subsection{Cosmic ray removal}

The \texttt{teareduce} package includes an auxiliary program called
\texttt{tea-cleanest}, which enables the interactive cleaning of cosmic rays in
individual exposures (see Fig.~\ref{fig3}). This code is inspired by the
\texttt{cleanest} code
\citep{Cardiel2020}\footnote{\url{https://cleanest.readthedocs.io/}}, although
the approach to detecting cosmic rays differs. In particular, this program uses
the L.A. Cosmic algorithm \citep{vanDokkum2001} to identify pixels suspected of
being affected by cosmic-ray hits. 

At the time of writing, other algorithms, such as PyCosmic
\citep{Husemann2012}, deepCR \citep{Zhang2020} and Cosmic-CoNN \citep{Xu2023},
are being incorporated into this tool.

\articlefigure[width=1.00\textwidth]{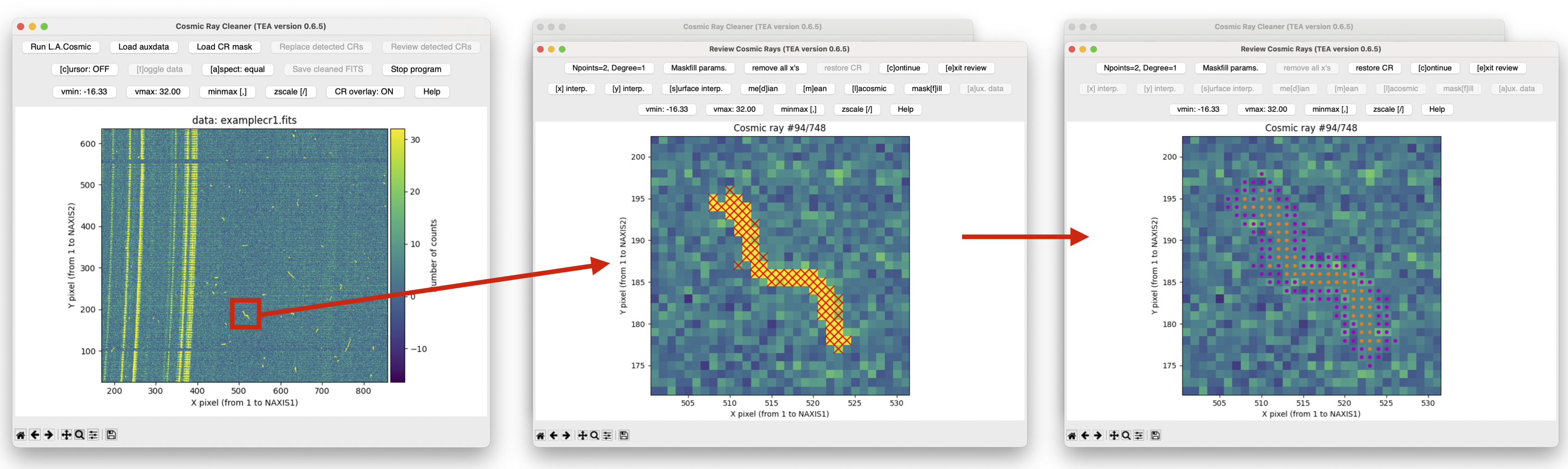}{fig3}{Example of cosmic ray
detection and correction: original image with multiple cosmic-ray hits
(\textit{left}), identification of a specific cosmic ray with pre-flagged
pixels marked by red x's allowing manual selection/deselection
(\textit{center}), and corrected image after interpolation with interpolated
pixels marked by x's and pixels used for interpolation indicated by magenta
dots (\textit{right}).}

\texttt{Teareduce} also provides an alternative method, the function
\texttt{cr2images}, for automatically removing cosmic rays in cases where two
equivalent exposures are available. In such situations, the cosmic-ray pixels
in each exposure can be replaced with the signal from the other exposure. This
function also allows the users to define regions to be cleaned or regions to be
excluded from the cleaning process.

\acknowledgements We thank the support of the projects PID2021-123417OB-I00 and
PID2022-138621NB-I00, funded by MCIN/AEI/10.13039/501100011033/FEDER, EU, and
PCI2022-135023-2, funded by MCIN/AEI/10.13039/501100011033 and by the European
Union 'NextGenerationEU'/PRTR.  This work made use of \texttt{astropy}
\citep{Astropy2013, Astropy2018}, \texttt{numpy} \citep{Harris2020},
\texttt{scipy} \citep{Virtanen2020}, \texttt{ccdproc} \citep{MattCraig2017},
and \texttt{matplotlib} \citep{Hunter2007}.

\bibliography{128}  


\end{document}